\documentstyle[a4]{article}

\def\ie{{\rm i.e.\/}\ }

\def\etc{{\rm etc\/}\ }

\def\ZZ{\mbox{\rm Z}\hskip-4pt \mbox{\rm Z}}
\def\RR{\mbox{\rm I}\hskip-2pt \mbox{\rm R}}

\def\CC{\mbox{\rm C}\hskip-6pt \mbox{l} \;}

\begin{document}
\def\ie{{\rm i.e.\/}\ }

\def\etc{{\rm etc\/}\ }

\begin{titlepage}

\begin{center}

\renewcommand{\thefootnote}{\fnsymbol{footnote}}

\centerline{{\Large Differentials of higher order}}
\smallskip
\centerline{{\Large in}}
\smallskip
\centerline{{\Large non commutative differential geometry}}
\vspace{1,5 cm}
\centerline{R. Coquereaux}

\date{October 16, 1996}

\vspace{1.cm}

{\it Centre de Physique Th\'eorique - CNRS - Luminy, Case
907}

{\it F-13288 Marseille Cedex 9 - France }

\end{center}

\begin{abstract}

In differential geometry, the notation $d^n f$ along
with the corresponding formalism has fallen into disuse since the birth of
exterior calculus. However, differentials of higher order are useful objects
that can be interpreted in terms of functions on iterated tangent bundles (or in term of jets).
We generalize this notion to the case of non commutative differential
geometry. For an arbitrary algebra ${\cal A}$, people already know how to
 define the differential algebra $\Omega{\cal A}$ of
 universal differential forms over ${\cal A}$.
We define Leibniz forms of order $n$ (these are {\em not} forms of degree $n$, \ie they are not
elements of  $\Omega^n{\cal A}$) as particular elements of what we call
the ``iterated frame algebra'' of order $n$, $F_n{\cal A}$, which is itself defined as the
$2^n$ tensor power of the algebra ${\cal A}$. We give a system of
generators for this iterated frame algebra and identify the ${\cal A}$ 
left-module
of forms of order $n$ as a particular vector subspace
included in the space of universal one-forms built over the iterated
frame algebra of order $n-1$. We study the algebraic
structure of these objects, recover the case of the commutative differential
calculus of order $n$ (Leibniz differentials) and give a few examples.
\end{abstract}

\vspace{3 cm}


\vspace{2 cm}

\noindent December 1996

\vspace{0.5 cm}

\noindent CPT-96/P.3403

\vspace{2 cm}

\noindent anonymous ftp or gopher: cpt.univ-mrs.fr

\vfill {}
\end{titlepage}
\newpage

{\Large \centerline{Differentials of higher order in non commutative differential
geometry}}

\section {Introduction}

Given a function $f$ on a manifold, one knows how to consider the collection of its first derivatives, with respect to a coordinate frame 
$\{\partial_\mu \doteq {\partial \over \partial x^\mu}\}$
as a ``global object'' : we just have to build  the one-form $df = \partial_\mu f dx^\mu$. When our manifold is riemannian, we may also use the metric
  $g = \{g^{\mu\nu}\}$, to ``rise'' the indices and build the gradient $grad \, f$, which is a vector field of
 components $g^{\mu\nu}\partial_\mu f $.  As everybody knows, the collection of first derivatives of $f$ transforms nicely (Leibniz rule!) under a change of coordinates. Things become more subtle at the level of second derivatives : here one usually needs a linear connection to build an invariant object.

 Differential forms of degree $2,3\ldots$ are very familiar objects that
 can be considered as fully
 antisymmetric tensor fields, but they do not provide the geometric objects
that one needs to handle  second or higher derivatives of a function 
(since $\partial_\mu \partial_\nu f = \partial_\nu \partial_\mu f$, the only two-form that we can build from second derivatives alone is equal to $0$).

As it is well known, a possibility is to introduce a linear connection $\nabla$, so that, if $\omega$ is a one-form,  we may consider the covariant derivatives (call $\omega_{\mu;\nu}$
the components of $\nabla \omega$). In particular, if $\omega = d f$, we may
consider the Hessian of $f$, namely $Hess \, f \doteq \nabla \nabla f = \nabla df$, which is a bilinear form on the tangent bundle (symmetric when there is no torsion). The components of $Hess \, f$, of course, transform nicely under a
change of coordinates, but they involve the linear connection. 

There is another possibility which does not require introduction of linear
connections and is based on the observation that
instead of the chain rule for first derivatives, $d\phi/dv = (d\phi/du) \, (du/dv)$, we have something
more complicated for second derivatives, namely,
$d^2\phi/dv^2 = (d^2\phi/du^2) \, (du/dv)^2 +( d\phi/du) \, d^2u/dv^2$.
Since transforming between coordinate systems requires that the second derivative be accompanied by the first, one may therefore try to put them together to form a new object. This idea lead mathematicians of the nineteenth century to introduce a formalism using ``differentials of higher orders'' (not to be confused with nowadays {\sl usual \/} differential forms of degree $p$) and notations such as $d^p f$.
This notation, along with the corresponding formalism has almost disappeared since the birth of exterior calculus (every student knows that $d^2 = 0$!) These differentials of higher orders can be actually be given a good invariant geometric status by working in the so called ``bundle of jets of infinite order'' or in iterated frame bundles.
The basic idea, that we illustrate in the case of forms of order $2$, is the following: let $f$ be a function on a manifold $M$, its differential
$df$ can be paired with a vector at a given point to give a number; $df$ is therefore a
function on the tangent bundle $TM$; but $TM$ is itself a manifold and one can look at the
differential $d_1 g$ ($d_1$ is not the same $d$ as before!) of a function $g$ on {\it this} manifold. In particular, one can build the object $\delta^2 f \doteq d_1 (d f)$ which is a function on $T^2M=T(TM)$. One can continue$\ldots$

There are, unfortunately, almost no papers trying to
fill the gap between the Leibniz formalism of differential forms of higher order and modern differential geometry (with the notable exception of \cite{Meyer}). We shall call these forms {\sl Leibniz forms of order\/} $n$ to distinguish them
from the usual (exterior) differential forms of degree ~$n$.

One motivation to resuscitate this almost forgotten formalism, in the context of usual (``commutative'') geometry comes from the theory of stochastic processes \cite{Schwartz} where it was shown that the natural objects that one should integrate along some irregular and continuous curves
(like brownian curves) are not the usual one-forms of differential geometry but$\ldots$ differentials of order $2$. Similar considerations could be made when considering fractal objects with integral Haussdorf dimension.

Our motivation, in the present paper, comes actually from quantum field theory and is two-fold.

Most constructions of quantum physics are usually expressed in terms of  operator algebras (or Feynmann graphs). We believe that it is useful to ``geometrize'' quantum field theory, \ie
to interpret its objects in terms of geometrical quantities like forms, connections \etc., but of course, this has to be done ``in a quantum way'', where non commutative algebras replace  the ordinary commutative algebras of smooth functions over manifolds.
To some extent, this attempt has been initiated and developed by several
people during the last few years (cf. in particular the works by A. Connes or M. Dubois-Violette). We believe that introducing a non commutative analogue for Leibniz forms of higher orders will turn out to be a useful step in this program.

Another (unrelated) physical motivation for our present work is the hope that
such a formalism could help people to devise consistent theories
involving fields with spin higher than $2$.

In a nutshell, 
the purpose of the present paper is to introduce, for any associative algebra ${\cal A}$ a formalism of differential
forms of higher orders that, in the case where ${\cal A}$ is the algebra of smooth functions on a differentiable manifold, specializes to the ``old'' Leibniz forms of order $n$ (we are {\sl not\/} trying  discuss a non commutative analogue for De Rham forms of degree $n$ since this is already well known$\ldots$)

To our knowledge, there are no other papers on the present subject in the literature (neither in mathematics nor in theoretical physics).

The forms of order $k$ (along with differentials satisfying relations
like $d^n k = 0$) introduced recently by \cite{MDV},
are not related to ours but to algebras deformations or to quantum groups.

\bigskip

\section{Algebras, universal differential algebras and iterated 
frame algebras}

Let ${\cal A}$ be a unital associative algebra over the field of complex numbers.
and call $1$ the unit of ${\cal A}$.

\begin{itemize}

\item

We call ${\cal T}^p {\cal A} \doteq {\cal A}^{\otimes (p+1)}$. Warning: 
there is a shift of the degree by $1$ so that ${\cal T}^0{\cal A}  = {\cal A}$. 

The vector spaces ${\cal T}^p {\cal A}$ are bi-modules over ${\cal A}$, 
with
\begin{eqnarray*}
a \times a_{0}\otimes a_{1}\otimes \ldots a_{n} & = & 
a a_{0}\otimes a_{1}\otimes \ldots a_{n}
\\
a_{0}\otimes a_{1}\otimes \ldots a_{n}  \otimes a & = & a_{0}\otimes a_{1}\otimes 
\ldots a_{n}a
\end{eqnarray*}

${\cal T} {\cal A} \doteq \bigoplus_{p}{\cal T}^p {\cal A}$ is a graded 
algebra with multiplication
$$
a_{0}\otimes a_{1}\otimes \ldots \otimes a_{p} \cdot b_{0}\otimes b_{1}\otimes 
\ldots \otimes b_{q} =
a_{0}\otimes a_{1}\otimes \ldots \otimes a_{p}b_{0}\otimes b_{1}\otimes 
\ldots \otimes b_{q} 
$$
We shall call ${\cal T} {\cal A}$ the ``${\cal T}$-algebra of ${\cal A}$''.
Notice that this is {\em not\/} the tensorial algebra over ${\cal A}$: 
the product of $a_{0}\otimes a_{1}$ and $b_{0}\otimes b_{1}$ 
in the tensorial algebra would be $a_{0}\otimes a_{1} \otimes b_{0}\otimes 
b_{1}$ whereas it is $a_{0}\otimes a_{1}b_{0}\otimes  b_{1}$ in ${\cal T} 
{\cal A}$.

\item

We denote by ${\cal A}_{p}$ the vector space ${\cal 
A}^{\otimes p}$ endowed with the product algebra structure inherited from ${\cal A}$.
For instance, ${\cal A}_{2} = {\cal A} \otimes {\cal A}$ has a 
multiplication defined by:
$(a_{1 }\otimes b_{1}) (a_{2}\otimes b_{2}) = a_{1 }a_{2}\otimes 
b_{1} b_{2}.$
Notice that, although ${\cal A}_{p+1}$ and ${\cal T}^p {\cal A}$ coincide, 
as vector spaces, their algebraic structures are totally 
different. The unit of the algebra ${\cal A}_{p}$ is $1_{p}\doteq 
1\otimes 1\otimes \ldots \otimes 1$, ($p$ times).

\item
Let  $m$ be the multiplication map of ${\cal A}$, \ie, $m(a \otimes 
b \in {\cal A} \otimes {\cal A}) = ab \in {\cal A}$.
Let $\Omega {\cal A}$ the universal differential algebra of ${\cal 
A}$. We recall its construction. $\Omega {\cal A} \doteq 
\bigoplus_{p}\Omega^p {\cal A}$ where $\Omega^{0}{\cal A} = {\cal A}$,
$\Omega^{1}{\cal A} = Ker(m) \subset {\cal T}^{1}{\cal A}$,  
$\Omega^{p}{\cal A} =  \Omega^{1}{\cal A} \otimes_{{\cal A}} \Omega^{1}{\cal A} \otimes _{{\cal A}} \ldots 
\Omega^{1}{\cal A} \subset {\cal T}^{p}{\cal A}$.
As usual, one defines $d_0 b \doteq 1 \otimes b - b \otimes 1$ so that 
$\Omega^{p}{\cal A}$ is the linear span of the monomials $a_{0}d_0 
a_{1}d_0 a_{2}\ldots d_0 a_{p}$ with the product inherited from ${\cal T} {\cal A}$.
Remember that  $\Omega {\cal A} \subset {\cal T} {\cal A}$.
When ${\cal A}$ is an algebra of functions over a space $M$ (hence a
commutative algebra), it is clear that
 we can identity $d_0 b$ with the function of two
variables  $[d_0 b](x,y)=b(y)-b(x)$.

\item
Let $m_{p}$ be the multiplication map in the algebra ${\cal A}_{p}$,  
\ie, 
$m_{p}(
(a_{1}\otimes a_{2} \ldots \otimes a_{p})
 \otimes 
(b_{1}\otimes b_{2} \ldots \otimes b_{p}) 
\in {\cal A}_{p} \otimes {\cal A}_{p} ={\cal A}_{2p})
=
a_{1}b_{1}\otimes a_{2}b_{2} \ldots \otimes a_{p}b_{p} \in {\cal 
A}_{p}$
We call $\Omega {\cal A}_{p}$ the universal differential algebra of ${\cal 
A}_{p}$ (and sometimes, the ``$\Omega$-algebra'' of ${\cal A}_p$). It is defined exactly as usual, but of course with the 
multiplication map $m_{p}$ and the unit $1_{p}$ of ${\cal A}_{p}$.
Its differential is called $d_{p-1}$. Notice that 
$\Omega {\cal A}_{p} = \bigoplus_{q=0}^{\infty}\Omega^q{\cal A}_{p}$ 
and that $\Omega^q {\cal A} \equiv \Omega^q{\cal A}_{1}$.

\item
We shall now define, for each $p$, a new object that we call ``the iterated 
frame algebra of order $p$ of ${\cal A}$''. We call
\begin{eqnarray*}
F_{{0}}{\cal A} & = & {\cal A} \\
F_{{1}}{\cal A} & = &{\cal A} \otimes {\cal A} = {\cal A}_{2}\\
F_{{2}}{\cal A} & = & 
{\cal A} \otimes {\cal A}\otimes {\cal A} \otimes {\cal A} = {\cal A}_{4}\\
F_{{p}}{\cal A} & = & 
F_{{p-1}}{\cal A} \otimes F_{{p-1}}{\cal A} = {\cal 
A}_{2^p}
\end{eqnarray*}
Therefore $F_{{p}}{\cal A}$ is nothing else than the algebra ${\cal 
A}_{2^p}$; its unit will be called ${\underline 1}_p \doteq 1_{2^p}$ and 
its multiplication map is ${\underline m}_p\doteq m_{2^p}$.
Its differential will be called ${\mathbf \delta}_p$, so that 
${\mathbf \delta}_p = d_{2^p-1}$:
$$ {\mathbf \delta}_p :  F_{{p}}{\cal A} \mapsto  F_{{p+1}}{\cal A} $$
Notice that $\delta_0 = d_0$.

\item
The basic idea underlying what follows is the fact that one can lift 
elements of ${\cal A}$ to any of the $F_{{p}}$ (see below) and also 
that $\Omega^{1} F_{  {p-1}}{\cal A}$ is a vector subspace of 
the algebra $F_{  {p}}{\cal A}$ 
so that differentials (of degree one) of appropriate lifts of elements 
of ${\cal A}$ can be considered themselves as elements of a new 
algebra on which one can operate again with {\em another\/} differential.

\end{itemize}

\section{Lifts}

Take $f \in {\cal A} = F_{  {0}}{\cal A}$. We extend it to 
all the $F_{  {p}}{\cal A}$  as follows: 
$f_{  {p}} = f_{  {p - 1}}\otimes {\underline 1}_{  
{p-1}}$. We also call $f_{{0}}=f$. Therefore, we have, for 
instance
\begin{eqnarray*}
f_{  {4}} & =  & f_{  {3}} \otimes {\underline 1}_{  {3}} 
\\
 & =  &  f_{  {2}} \otimes {\underline 1}_{  {2}} \otimes {\underline 1}_{  {3}}
\\
 & =  &  f_{  {1}} \otimes {\underline 1}_{  {1}} \otimes {\underline 1}_{  {2}} 
 \otimes {\underline 1}_{  {3}}                    
\\
& =  &  f \otimes 1 \otimes {\underline 1}_{  {1}} \otimes {\underline 1}_{  {2}} 
\otimes {\underline 1}_{  {3}}   = f \otimes 1_{1+2+2^{2}+2^{3}=15} \in  
F_{  {4}}{\cal A} =  {\cal A}_{16}
\end{eqnarray*}
It is clear that $f_p=f_0 \otimes {\underline 1}_{2^p - 1}$
Ultimately we shall write $f$ for $f_{{p}}$, for any $p$. 
This useful  abuse of notations amounts, in commutative differential geometry, to 
identify a function on a manifold with its pullback in the tangent bundle 
and, actually, with all its pullbacks in the tower of iterated 
tangent bundles.

Let us call $L_{  {p}}{\cal A}$ the algebra of lifts from ${\cal A}$ to
 $F_{{p}}{\cal A}$: these are tensors with one element of ${\cal 
 A}$ in the first position, followed by a string of tensor products with $1$'s
 (namely $2^p - 1$ of them). Clearly $L_{  {p}}{\cal A}$ is 
 an algebra isomorphic with ${\cal A}$.
 
Not only can we lift $f$ from $F_{  {0}}{\cal A}={\cal A}$ to $F_{  {p}}{\cal 
A}$ by the above trick, but also lift any $\omega \in F_{  {q}}{\cal 
A}$ to all the $F_{  {p}}{\cal A}$, $(p \, > \, q)$ by the 
same method: call $\omega_{{q}} \doteq \omega$,
 $\omega_{{q+1}} \doteq \omega_{{q}} \otimes 
 {\underline 1}_{{q}}$, $\ldots$, $\omega_{{p}} \doteq 
 \omega_{{p-1}} \otimes 
 {\underline 1}_{{p-1}}$.
Clearly, there are many elements of $F_p{\cal A}$ that 
 are not lifts (neither lifts from ${\cal A}$, nor lifts from
 $F_{   {q}}{\cal A}$, with $q \, < \, p$).
 
 The ``elementary'' lifting operation itself from $F_{  {s}}{\cal A}$  to 
$ F_{  {s+1}}{\cal A}$  will be denoted by $\rho_{s}$ (so 
 that the index $s$ of $\rho_s$ reminds us where we start from).
 For instance, if $\alpha \in F_{  {s}}{\cal A}$, then
 $\rho_{s} \alpha \doteq \alpha \otimes {\underline 1}_{{s}} \in  F_{  {s+1}}{\cal A}$.

 Notation: we shall call $\omega_{{p}}$ the lift of 
 $\omega$  to $F_{  {p}}{\cal A}$, {\em wherever $\omega$ is located\/}
 in the tower of algebras $F_s{\cal A}$'s,
 when we want to remember in which such algebra
 we are$\ldots$ but it is often convenient to make a notational 
 abuse and write $\omega$ for $\omega_{\underline {p}}$, for any $p$.
 
  We shall say that $u$ is a $p$-element if it belongs to $F_{  
 {p}}{\cal A}$. Since $L_{  {p}}{\cal A} \subset F_{  
 {p}}{\cal A}$, it is clear that every element of ${\cal A}$ defines 
 infinitely many $p$-elements (one for each $p$), namely, the 
 collection of its lifts.
 
We have defined the lifting operator $\rho_s$ (right multiplication
by ${\underline 1}_s$) but we shall, at times, also need to use
the map  $\lambda_s$ (left multiplication by ${\underline 1}_s$) :
if $\alpha \in F_{  {s}}{\cal A}$, then
 $\lambda_{s} \alpha \doteq  {\underline 1}_{{s}}\otimes \alpha \in  F_{  {s+1}}{\cal A}$.
 
\section{Usual differentials and their lift in the tower of  $F_{  {p}}{\cal A}$}

Take $f \in F_{   {0}}{\cal A}={\cal A}$ and consider its 
usual differential $\delta_{0}f = 1 \otimes f - f \otimes 1$ in the universal
differential algebra $\Omega {\cal A}$. We may now consider  
$\delta_{0}f$ as an element of $F_{  {1}}{\cal A} = {\cal A} 
\otimes {\cal A}$ and lift it to the whole tower of $F_{  
{p}}{\cal A}$, for instance 
$(\delta_{0}f)_{3} = \rho_{2} \rho_{1} \delta_{0}f \in F_{  {3}}{\cal A}$.
Of course, we have $(\delta_{0}f)_{1} \equiv \delta_{0}f$.

There is no reason why we should stop at the level of $F_{  
{1}}{\cal A}$ and we shall also consider one-forms belonging to the
 $\Omega$-algebra of $F_{  
{2}}{\cal A}$, $F_{  {3}}{\cal A}$, \etc along with their 
own lifts. 
Actually, in the present paper, we only need to consider 
{\em usual one-forms and their lifts\/}, more precisely,
 the elements of $\Omega^{1} F_{  {p}}{\cal A}$ 
that we identify with elements of 
$F_{  {p+1}}{\cal A} = F_{  {p}}{\cal A}\otimes F_{  {p}}{\cal A}$ 
and that we subsequently lift to the whole tower of iterated frame algebras.

Notice that, for any $p$, the usual differential $d_{p-1}$ acting 
in the differential algebra $\Omega({\cal A}_{p}) = \bigoplus_{q} \Omega^q ({\cal A}_{p})$ is such 
that $d_{p-1}^{2}=0$, as usual.
In particular,
 the usual differential $\delta_{p-1}$ acting 
in the differential algebra $\Omega({\cal F}_{p}{\cal A}) =
 \bigoplus_{q} \Omega^q ({\cal F}_{p}{\cal A})$ is such 
that $\delta_{p-1}^{2}=0$.

 However, it should be clear that 
$\delta_{p+1}\delta_{p}$, for instance, or more generally $d_{p+1}d_{p}$,
 is {\em not \/} zero.
 Take $f \in F_{   {0}}{\cal A}={\cal A}$, we may build
$\delta_{0}f \in \Omega^{1} {\cal A} \subset  F_{   {1}}{\cal A} $.
 We can then apply the differential $\delta_{1}$ and obtain
$\delta_{1}\delta_{0}f \in \Omega^{1} F_{  {1}}{\cal A} \subset
 F_{   {2}}{\cal A} $.
 We can then again apply the differential $\delta_{2}$ and obtain
$\delta_{2}\delta_{1}\delta_{0}f \in \Omega^{1} F_{  {2}}{\cal A} \subset
 F_{   {3}}{\cal A} $.
 In this way, we build $\delta^p f \doteq \delta_{p-1}\delta_{p-2}\ldots \delta_{2}\delta_{1}\delta_{0}f
 \in F_{  {p}}{\cal A}$. This is our first example of a ``form 
 of order $p$'' and introduces, at the same time, the notation $\delta^p f$, for $f$, an 
 element of ${\cal A} $.
 
As usual, $\delta_{0}^2 = \delta_{1}^2 = \ldots =
 \delta_{p}^2 = 0$ but $\delta^{2} \neq 0$ and, more generally   
 $\delta^{p} \neq 0$. We can also lift the quantity $\delta^p f$, an 
 element of $F_{  {p}}{\cal A}$ to 
 any $F_{  {q}}{\cal A}$, with $q \, > \, p$ by using right 
 multiplication by the appropriate tensorial power of the unity.

 Let us conclude this subsection by stressing that $\Omega^{1} F_{  
 {p-1}}{\cal A} \subset
 F_{   {p}}{\cal A} $ so that $\delta_{p-1}$ maps 
  $F_{   {p-1}}{\cal A}$ to $ F_{   {p}}{\cal A}$.
  
\section{A set of generators for the iterated frame algebras over ${\cal A}$}

\subsection{The generators $\delta_I$}

Since $F_{   {p}}{\cal A}$ is a product of $2^p$ tensorial 
powers of ${\cal A}$, it is natural to define generators 
$\delta_{I}f$, $f\in {\cal A}$, parametrized by the set of subsets of 
a set with $p$ elements. Call $E_{p}=\{p-1, p-2, \ldots, 2,1,0\}$ and let 
$I \subset E_{p}$. In order to specify in a unique way the writing 
of $I$, we order the set in a decreasing way. Let us define 
$\delta_{I}$ by using an example. We want to find a set of generators 
in $F_{  {5}}{\cal A}$. Choosing, for instance $I = 
\{3,1,0\}\subset E_{5}$, we define 
$$\delta_{\{3,1,0\}} f \doteq  \rho_{4}\, \delta_{3}\, \rho_{2}\, \delta_{1}\, \delta_{0}\, f$$
Occurence of the needed $\delta_{p}$ is specified by the 
notation itself (here $\delta_{\{3,1,0\}}$) and the appropriate lifts (the
$\rho_{s}$) are associated with the complement of $I$ in $E_{p}$. 
Notice that when $A$ is the commutative algebra of functions on a 
manifold, the quantity $\delta_{\{3,1,0\}}f$ is a function of $2^{5}=32$ 
variables. From the definition of the lift operation $\rho_{s}$ 
and of the differential $\delta_{s}$, it is easy to see that the set 
of $\delta_{I}f$, when $I$ runs in the set of subsets of $E_{p}$, and 
when $f \in {\cal A}$ is indeed a family of generators for $F_{   
{p}}{\cal A}$.

 It is instructive to work out explicitly, in terms of 
tensor products, such a set of generators for $F_{   {2}}{\cal A}$.
\begin{eqnarray*}
\delta_{\emptyset}f & = & \rho_{1}\rho_{0} f  \\
\delta_{\{0\}}f & = & \rho_{1}\delta_{0} f  \\
\delta_{\{1\}}f & = & \delta_{1}\rho_{0} f  \\
\delta_{\{1,0\}}f & = & \delta_{1}\delta_{0} f  \\
\end{eqnarray*}
This reads, explicitly
\begin{eqnarray*}
\delta_{\emptyset}f & = & f\otimes 1 \otimes 1 \otimes 1 \\
\delta_{\{0\}}f & = &  1\otimes f \otimes 1 \otimes 1 - f \otimes 1 \otimes 1 \otimes 1 \\
\delta_{\{1\}}f & = & 1\otimes 1\otimes f\otimes 1 - f \otimes 1\otimes 1 \otimes 1  \\
\delta_{\{1,0\}}f & = &  1 \otimes 1 \otimes 1 \otimes f - 
1 \otimes 1 \otimes f \otimes 1 - 1 \otimes f \otimes 1 \otimes 1
+ f \otimes 1 \otimes 1 \otimes 1 \\
\end{eqnarray*}
We see, on this example, that the above $4$ types of monomials indeed span
$F_{{2}}{\cal A}$ when $f$ runs in ${\cal A}$ since we can invert this
system to obtain
\begin{eqnarray*}
f \otimes 1 \otimes 1 \otimes 1 & = & \delta_{\emptyset} f \\
1 \otimes f \otimes 1 \otimes 1 & = & \delta_{\{0\}} f  +  \delta_{\emptyset}f \\
1 \otimes 1 \otimes f \otimes 1 & = & \delta_{\{1\}} f  +  \delta_{\emptyset}f \\
1 \otimes 1 \otimes 1 \otimes f & = & \delta_{\{1,0\}}f + \delta_{\{1\}} f
 \delta_{\{0\}} f  +  \delta_{\emptyset}f \\
\end{eqnarray*}
 It is also instructive to work out explicitly, in terms of 
tensor products, such a set of generators for $F_{   {3}}{\cal A}$.
\begin{eqnarray*}
\delta_{\emptyset}f & = & \rho_{2}\rho_{1}\rho_{0} f  \\
\delta_{\{0\}}f & = & \rho_{2}\rho_{1}\delta_{0} f  \\
\delta_{\{1\}}f & = & \rho_{2}\delta_{1}\rho_{0} f  \\
\delta_{\{2\}}f & = & \delta_{2}\rho_{1}\rho_{0} f  \\
\delta_{\{1,0\}}f & = & \rho_{2}\delta_{1}\delta_{0} f  \\
\delta_{\{2,0\}}f & = & \delta_{2}\rho_{1}\delta_{0} f  \\
\delta_{\{2,1\}}f & = & \delta_{2}\delta_{1}\rho_{0} f  \\
\delta_{\{2,1,0\}}f & = & \delta_{2}\delta_{1}\delta_{0} f 
\end{eqnarray*}
We shall leave to the reader  the task of calculating the explicit expression
of these generators, in terms of tensor products.
Again, one can  see, on this example, that the above $8$ types of monomials indeed span
$F_{   {3}}{\cal A}$ when $f$ runs in ${\cal A}$.

Another abuse of notation: We shall often remove totally the symbol 
$\rho_{n}$. Indeed, it should be clear, 
when working in $F_3{\cal A}$, for example, that
$\delta_{2} \delta_{0} f$ denotes actually
$\delta_{\{2,0\}}f =\delta_{2}\rho_{1}\delta_{0} f$ since one needs to lift 
$\delta_{0} f$ to $F_{   {2}}{\cal A}$ in order to be able 
to use the differential $\delta_{2}$. Also, we shall write $\delta_p$ rather
than $\delta_{\{p\}}$ and even $\delta_{310}$ for $\delta_{\{3,1,0\}}$,
for example, since no confusion should arise.

\subsection{Right and left module multiplications}

Our purpose, in this subsection, is to warn the reader against a possible
subtle mistake.

For any associative algebra ${\cal B}$, the vector space $\Omega^1{\cal B}$ is
both a left and right module over  ${\cal B}$. This means in particular that,
if $a$ is an arbitrary element in ${\cal B}$
 and if $\omega$ is an arbitrary element in  $\Omega^1{\cal B}$, we have
$a \times \omega = (a \otimes 1) (\omega)$ where the multiplication involved
on the right hand side is the multiplication in the algebra 
 ${\cal B}\otimes {\cal B}$ whereas $\times$ denotes the left
module multiplication of elements of  $\Omega^1{\cal B}$ by elements of
 ${\cal B}$.
In the same way we have
$ \omega \times a = (\omega)  (1\otimes a)$. Notice that, in the first case, we
lifted $a$ from  ${\cal B}$ to  ${\cal B} \otimes {\cal B}$ by multiplying
it tensorially from the right by the unit, whereas, in the second case, we
had to perform this tensorial multiplication from the left.

Now, if we choose ${\cal B} = {\cal F}_p{\cal A}$, we see that
$\Omega^1{\cal F}_p{\cal A} \subset {\cal F}_{p+1}{\cal A}$ is both a
left and right module over the algebra ${\cal F}_p{\cal A}$. The above
remarks are of course valid.

To appreciate the subtlety, let us continue to denote explicitly
by $\times$ the module multiplication and
let us take $f$ and $g$ in ${\cal A}$, so that
$g \times \delta_0 f \in \Omega^1 {\cal A} \subset {\cal F}_{1}{\cal A}$. We can then
consider $\delta_1 (g \times \delta_0 f)$ which is an element of
$ \Omega^1 {\cal F}_{1}{\cal A} \subset {\cal F}_{2}{\cal A}$, hence a right
and left module over $ {\cal F}_{1}{\cal A}$ . The Leibniz rule being
true, we have certainly
$$
\delta_1 (g \times \delta_0 f) = (\delta_1 g) \times (\delta_0 f) +
g \times \delta_1(\delta_0 f)
$$
In order to check this relation, let us compute separately the left
and right hand sides of this relation. 

The term appearing on the r.h.s. is easy to calculate.
$$g\times \delta_0 h = (g\otimes 1)(1\otimes h - h \otimes 1)=
g\otimes h - gh \otimes 1
$$
Therefore,
$$
\delta_1(g\times \delta_0 h) = {\underline 1}_1 \otimes (g\times \delta_0 h)
- (g\times \delta_0 h) \otimes  {\underline 1}_1 =
1\otimes 1 \otimes g \otimes h -
1 \otimes 1 \otimes gh \otimes 1 -
g \otimes h \otimes 1 \otimes 1
+ gh \otimes 1 \otimes 1 \otimes 1
$$
and
therefore,
$$
\delta_1(\delta_0 h) = {\underline 1}_1 \otimes (\delta_0 h)
- (\delta_0 h) \otimes  {\underline 1}_1 =
1\otimes 1 \otimes 1 \otimes h 
- 1 \otimes 1 \otimes h \otimes 1
- 1 \otimes h \otimes 1 \otimes 1
+ h \otimes 1 \otimes 1 \otimes 1
$$
The last term of the r.h.s. is therefore also easy to calculate :

$$g \times \delta_1 \delta_0 h = g \otimes 1 \otimes 1 \otimes h
- g  \otimes 1  \otimes h  \otimes 1 - g  \otimes h  \otimes 1  \otimes 1 +
gh  \otimes 1  \otimes 1  \otimes 1
$$
One has however to be cautious with the first term of the r.h.s, indeed, it reads
$$
(\delta_1 g)\times (\delta_0 h) = (\delta_1 (g \otimes 1)) \times
(1\otimes h - h \otimes 1)$$
Replacing $g$ by $g\otimes 1$ in the above should be clear since we have to
lift ${\cal A}$ to ${F}_1{\cal A}$, so that
$(\delta_1 (g \otimes 1))={\underline 1}_1 \otimes (g\otimes 1) -
(g\otimes 1)\otimes {\underline 1}_1$.
What may lead to a possible mistake is the evaluation of the module
 multiplication $\times$ since we have here to take 
 ${F}_2{\cal A}$ as a {\it right\/} module over  ${ F}_1{\cal A}$.
In other words, one trades the module right multiplication ($\times$) with the
algebra multiplication in ${\cal A}^{\otimes 4}$
(please, notice the appearance of the {\em left} multiplication by ${\underline 1}_1$) in such a way that
$$
({\underline 1}_1 \otimes (g\otimes 1) -
(g\otimes 1)\otimes {\underline 1}_1)\times
 (1\otimes h - h \otimes 1) =
({\underline 1}_1 \otimes (g\otimes 1) -
(g\otimes 1)\otimes {\underline 1}_1)
({\underline 1}_1 \otimes (1\otimes h - h \otimes 1))
$$
Finally,
$$
(\delta_1 g)\times (\delta_0 h) = 
1 \otimes 1  \otimes g  \otimes h -
g  \otimes 1  \otimes 1  \otimes h -
1  \otimes 1  \otimes gh  \otimes 1 +
g  \otimes 1  \otimes h  \otimes 1
$$
adding the two contributions we see that the Leibniz rule is 
satisfied, as it should.

In section 3, we defined the lifting operator $\rho_s$ (right multiplication
by ${\underline 1}_s$) but we also defined the map
 $\lambda_s$ (left multiplication by ${\underline 1}_s$).
Using both maps $\rho_s$ and $\lambda_s$, we see that the above Leibniz rule can be written
$$
\delta_1 (g \times \delta_0 h) = (\delta_1 g) \times (\delta_0 h) +
g \times \delta_1(\delta_0 h)
$$
but also
$$
\delta_1((\rho_0 g)(\delta_0 h)) = (\delta_1 \rho_0 g) (\lambda_1 \delta_0 h) +
(\rho_1 \rho_0 g) (\delta_1(\delta_0 h))
$$
The reader can also check that
$$
\lambda_1 \delta_0 h = \delta_{\{1,0\}} h + \delta_{\{0\}} h
= \delta_1\delta_0 h + \rho_1 \delta_0 h
$$

\section{Higher differentials in non commutative geometry}
For the reader directly jumping to this section, let us repeat that we are not
trying to define the analogue of De Rham $n$-forms (forms of {\em degree} $n$)
 in non commutative differential geometry, for the good reason that it has been done and 
that it is well known. The ``differentials of higher order''
 of our title refer 
to objects whose commutative analogue are the quantities already
 called $d^n f$ by mathematicians from the previous century and that one 
can interpret, nowadays, in terms of iterated tangent bundles. From the 
terminological point of view, we shall therefore distinguish the 
(usual)``$n-forms$'' \ie forms of degree $n$ from the ``Leibniz forms of order $n$''.

 We already met one 
kind of {\em differential forms of order $n$}, namely, elements of the kind
 $\delta^n f = 
\delta_{n-1}\ldots \delta_1\delta_0 f$. 
We define inductively the space ${{\cal D}_{n}{\cal A}}$ of forms of order $n$ 
as follows.
$$ {{\cal D}_{n}{\cal A}} \doteq L_{n}{\cal A} \, \times \,\delta_{n-1} {{\cal D}_{n-1}{\cal A}}$$
The induction starts by taking 
${{\cal D}_{1}{\cal A}} \doteq \Omega^{1}{\cal A} = {\cal A}\, \times \, \delta_{0}{\cal A} 
\subset F_{   {1}}{\cal A}$

In other words, since  $L_{n}{\cal A}$ is isomorphic with ${\cal A}$,
${{\cal D}_{n}{\cal A}}$ is a left-module over ${\cal A}$
generated by the differentials $\delta_{n-1}\omega$, where 
$\omega \in {{\cal D}_{n-1}{\cal A}}$.

\section{Identification of ${{\cal D}_{n}{\cal A}}$ within $F_{  {n}}{\cal A}$}

Since ${{\cal D}_{n}{\cal A}}\subset F_{  {n}}{\cal A}$, it is a 
priori possible to express the differential forms of order $n$ in 
terms of the generators $\delta_{I}f$, where $I \subset E_{n}$.
\begin{itemize}

\item The case ${{\cal D}_{1}{\cal A}}$. Generators of $F_{  {1}}{\cal 
A}$ are of the kind $\delta_{\emptyset}f = f$ and $\delta_{\{0\}}f = 
\delta_{0}f$ so that forms of order one are just the ``usual'' 
universal one-forms (universal forms of degree one)
, \ie ${{\cal D}_{1}{\cal A}} = \Omega^{1}{\cal A}$. With 
$f,g \in {\cal A}$, they can be written as linear combinations of $f
\delta_{0}g$.

\item The case ${{\cal D}_{2}{\cal A}}$. Generators of $F_{  {2}}{\cal 
A}$ are of the kind $\delta_{\emptyset}f = f$,  $\delta_{\{0\}}f = 
\rho_{1}\delta_{0}f$, $\delta_{\{1\}}f = \delta_{1}\rho_{0}f$ 
and $\delta_{\{1,0\}}f = \delta_{1}\delta_{0}f$. By definition,
 forms of order $2$ 
are obtained by taking the $\delta_{1}$ of forms of order $1$ and by 
left-multiplying with arbitrary elements in  ${\cal A}$. This left 
multiplication is the external module multiplication, but one can, as 
well, lift ${\cal A}$ to ${ L_{ {2}}{\cal A}} \subset F_{  {2}}{\cal 
A}$ and use the algebra multiplication in the latter; the result is of 
course the same. 
 With $f,g,h \in {\cal A}$, forms of order two are linear combinations of terms
of the kind $f\delta_1(g\delta_0 h)$, and we have seen, in section 5, that,
explicitly (in terms of tensor products), such forms read
$ f(\delta_{1}\rho_{0}g) (\lambda_{1}\delta_{0}h) + 
fg \delta_{1}\delta_{0}h $.
 It is convenient to introduce indexed 
symbols  $a_{\mu}$, $a_{\mu \nu}$ and $x^\mu$ 
 refering to arbitrary elements of ${\cal A}$. When this algebra is 
 commutative (hence ${\cal A} = C(M)$ for some topological compact 
 space $M$) we can of course interpret these elements as functions on $M$.
 Expanding over the generators of $F_{  {2}}{\cal 
A}$, we see that a generic element of ${{\cal D}_{2}{\cal A}}$ can be written as
$$
\omega  =   a_{\mu} \delta_1 \delta_0 x^\mu +  a_{\mu \nu} \delta_{1} x^\mu \times \delta_{0} x^\nu
$$
Here, for the last time, we explicitly use the symbol $\times$ to distinguish the
module multiplication from the algebra multiplication$\ldots$
Moreover, both kinds of terms appearing on the right hand side
separately belong to ${\cal D}_2{\cal A}$. Indeed, from the definition, it is clear that 
$  a_{\mu} \delta_1 \delta_0 x^\mu$ is a differential form of order $2$; moreover, we have
   $\delta_{1}x^\mu \delta_{0}x^\nu = 
 \delta_{1}(x^\mu \delta_{0}x^\nu) - x^\mu \delta_{1}\delta_{0} 
 x^\nu$, which is the difference of two terms already shown to belong 
 to ${{\cal D}_{2}{\cal A}}$, so that
 $\delta_{1}x^\mu \delta_{0}x^\nu \in D_{2}{\cal A}$ as well.

We shall introduce later a new associative 
 product (and a new symbol) $\odot$, that will allow us to rewrite 
 the above as 
 $$\omega =  a_{\mu} \delta^{2} x^\mu +
 a_{\mu \nu} \delta x^\mu \odot  \delta x^\nu$$
 remember that the symbol $\delta^{2}f$ was already introduced to denote 
 $\delta_{\{1,0\}}f = \delta_1 \delta_0 f$.

\item The case ${{\cal D}_{3}{\cal A}}$.
Generators of $F_{{3}}{\cal A}$ are of the kind
 $\delta_{\emptyset}f = f$,  $\delta_{\{0\}}f$,
 $\delta_{\{1\}}f$, $\delta_{\{2\}}f$, 
 $\delta_{\{1,0\}}f$, $\delta_{\{2,0\}}f$, $\delta_{\{2,1\}}f$
  and $\delta_{\{2,1,0\}}f$.
  Forms of order $3$ 
are obtained by taking the $\delta_{2}$ of forms of order $2$ and by 
left-multiplying with arbitrary elements in  ${\cal A}$.

It is clear that $\delta_{\{2,1,0\}}f \in {{\cal D}_{3}{\cal A}}$.

Then $\delta_{2}x^\mu \delta_{\{1,0\}}x^\nu =
 \delta_{2}(x^\mu \delta_{\{1,0\}}x^\nu) - x^\mu 
 \delta_{\{2,1,0\}}x^\nu \in {{\cal D}_{3}{\cal A}}$ as a difference of two 
 elements of ${{\cal D}_{3}{\cal A}}$. Hence 
 $b_{\mu\nu}\delta_{2}x^\mu \delta_{\{1,0\}} x^\nu \in {{\cal D}_{3}{\cal A}}$.
 
 Also $\delta_{2}(\delta_{1}x^\mu \delta_{0} x^\nu)= 
 \delta_{\{2,1\}}x^\mu \delta_{0}x^\nu + 
 \delta_{1}x^\mu\delta_{\{2,0\}}x^\nu$, so that 
 $a_{\mu\nu}(\delta_{\{2,1\}}x^\mu \delta_{0}x^\nu + 
 \delta_{1}x^\mu\delta_{\{2,0\}}x^\nu)\in {{\cal D}_{3}{\cal A}}$.
 
 Finally $\delta_{2}x^\mu\delta_{1}x^\nu\delta_{0}x^\lambda = 
 \delta_{2}(x^\mu\delta_{1}x^\nu\delta_{0}x^\lambda)-
 x^\mu (\delta_{\{2,1\}}x^\mu \delta_{0}x^\nu + 
 \delta_{1}x^\mu\delta_{\{2,0\}}x^\nu)$
 therefore, 
 $a_{\mu\nu\lambda}\delta_{2}x^\mu\delta_{1}x^\nu\delta_{0}x^\lambda\in {{\cal D}_{3}{\cal A}}.$

A generic element of ${{\cal D}_{3}{\cal A}}$ can therefore be written as 

\begin{eqnarray*}
\omega  & = & 
a_{\mu}\delta_{\{2,1,0\}}x^\mu + 
b_{\mu\nu}\delta_{2}x^\mu\delta_{\{1,0\}}x^\nu + 
a_{\mu\nu}(\delta_{\{2,1\}}x^\mu \delta_{0}x^\nu +
   \delta_{1}x^\mu\delta_{\{2,0\}}x^\nu) +  \\ 
{} & {} & a_{\mu\nu\lambda}\delta_{2}x^\mu\delta_{1}x^\nu\delta_{0}x^\lambda
\end{eqnarray*}

\item The general case ${{\cal D}_{s}{\cal A}}$. It is already clear
from the study of the previous examples that inclusion of ${{\cal D}_{s}{\cal A}}$ in
 $F_s{\cal A}$ is strict; moreover, it is easy to see that, as a ${\cal A}$-module,
the former is of rank $2^{s-1}$ whereas the latter is of rank $2^s$.
Writing differential forms of order $s$ in terms of generators of
$F_{s}{\cal A}$, although straightforward, is not always very illuminating,
as one can infer from the previous examples. As we shall see, it is much better
to introduce the ``old'' Leibniz notation together with a new product $\odot$.

\end{itemize}

\section{The graded differential algebra ${{\cal D}, \odot, \delta}$}

We want to define on ${{\cal D}}$ a structure of an associative (non 
commutative) algebra (\ie a product $\odot$), for which $\delta$
will be a derivation mapping forms of 
order $n$ to forms of order $n+1$. Warning: in order not to confuse 
the reader, we shall not call this structure a ``differential algebra'', 
first because $\delta$ is not of square zero (none of its powers is, {\it 
a priori}, equal to zero), next because, although  ${{\cal D}}$ is 
$\ZZ$-graded and $\delta$ will be a derivation of ${{\cal D}}$, it will not be 
a graded derivation.
We first explain how to construct the new product and then
show how to re-express all forms of order $n$ with this notation. 
The obtained algebra $({{\cal D}}, \odot)$ is a bimodule over ${\cal A}$,
but left and right multiplications by elements of ${\cal A}$ do not coincide, so that
${\cal A}$ is an algebra over $\CC$ only. This is {\it a priori} clear in non commutative
geometry, \ie, when ${\cal A}$ is not commutative but these two operations do not
coincide even when  ${\cal A}$ is commutative may be a surprise...

\subsection{The product $\odot$}
Here --- and in general --- latin letters like $f,g,h\ldots$ refer to elements of the
algebra ${\cal A}$. Moreover, we decide ro write $\delta \omega$ rather than $\delta_q \omega$
when $\omega$ is a form of order $q$.

\begin{description}
	\item[Products of forms of order $0$ by arbitrary forms of order $q$.]

	Take $f \in {\cal A}$  and 
$\sigma$ an arbitrary form of order $q$, \ie
 $\sigma \in {{\cal D}_{q}{\cal A}}$.
The (left) product with an element of ${\cal A}$ is the usual product:
$$ f \odot \sigma = f \sigma$$

	\item[Products of forms of order $1$ by arbitrary forms of order $q$.]

	Next we define the product of a one-form of the kind $\delta f$
 with an 
arbitrary $\sigma$. Notice that, in what follows,
 the right hand side is already defined:

$$
\delta f \odot \sigma \doteq \delta (f \sigma) - f \delta \sigma
$$

An arbitrary form of order $1$ can be written as $f \delta g$. One defines 
the product
$$
(f \delta g) \odot \sigma \doteq  f (\delta g \odot \sigma)
$$

	\item[Products of forms of order $2$ by arbitrary forms of order $q$.]

	 We now define the product of a form of order two, times an arbitrary 
form of order $q$.
There are two kinds of forms of order two: those of the kind $\delta f \odot \delta g$
 and those of the kind $\delta^2 f$. We define the products as
 
 $$ (\delta f \odot \delta g) \odot \sigma \doteq \delta f \odot (\delta g \odot \sigma)$$
 and
 $$ \delta^2 f \odot \sigma \doteq \delta(\delta f \odot \sigma) - \delta f \odot \delta \sigma$$

	\item[Generalisation]

  The general procedure should be clear. Using recusion on $p$,
we can define the product $\odot$ of a form of order 
	$p$ times an arbitrary form $\omega$, by imposing 
	associativity and the Leibniz rule. The only thing to check is that 
	this definition
is indeed compatible with associativity of $\odot$. We leave 
	this task to the reader.
	
\end{description}

\subsection{Forms of order $1,2,3,4,\ldots$}

Using the $\odot$ product, one can write all the higher differential
forms of order $n$ in a very simple way since it allows us to use the
``old'' Leibniz notation (as in the commutative case, see \cite{Meyer}).
Again, in what follow, letters $f,g,h,i,k \ldots$ refer to elements in ${\cal A}$.
As an example, we give the structure of all possible types of differential
forms of order $0,1,2,3,4$ and express each of them in terms of the
differential operators $\delta_p$ and more generally, in terms of the
generators $\delta_I$. We not write
explicitly the lifts $\rho_s$ (or $\lambda_s$).
\begin{description}
\item[Forms of order 1.] Elements of ${\cal D}_1{\cal A}$ are linear
combinations of $f\delta g$ where
$$\delta g = \delta_0 g$$
\item[Forms of order 2.]  Elements of ${\cal D}_2{\cal A}$ are linear
combinations of two types: $f\delta^2 g$ and $f\delta g \odot \delta h$, where
\begin{eqnarray*}
\delta^2 g & = & \delta_1\delta_0 g \\
\delta g \odot \delta h & = & \delta_1 g \delta_0 h 
\end{eqnarray*}
\item[Forms of order 3.]Elements of ${\cal D}_3{\cal A}$ are linear
combinations of four types: $f\delta^3 g$, $f(\delta g\odot\delta^2 h)$,
$f(\delta^2 g \odot \delta h)$, and $f(\delta g\odot\delta h\odot\delta i)$, where
\begin{eqnarray*}
\delta^3 g & = & \delta_2 \delta_1 \delta_0 g \\
\delta g \odot \delta^2 h  & = & \delta_2 g \delta_{10} h \\
\delta^2 g \odot \delta h  & = & \delta_{21}g\delta_0 h +
 \delta_1 g \delta_{20}h - \delta_2 g \delta_{10} h \\
\delta g \odot \delta h \odot \delta i & = & \delta_2 g \delta_1 h \delta_0 i
\end{eqnarray*}
The only non obvious kind of terms is the third one; it is obtained as a difference, as explained before,  by imposing the Leibniz rule.
\begin{eqnarray*}
\delta^2 g \odot \delta h
& = &  \delta(\delta g \odot \delta h) - \delta g \odot \delta^2 h \\
& = &  \delta_2(\delta_1g \delta_0 h) - \delta_2 g \delta_{10} h \\
& = &  \delta_2 \delta_1 g \delta_0 h + \delta_1 g \delta_{20} h -
 \delta_2 g \delta_{10} h
\end{eqnarray*}

\item[Forms of order 4.]
Elements of ${\cal D}_4{\cal A}$ are linear
combinations of eight types:
$k \delta^4 f$,
$k( \delta f \odot \delta^3 g)$,
$k( \delta f \odot \delta^2 g \odot \delta h)$,
$k( \delta f \odot \delta g  \odot \delta^2 h)$,
$k( \delta f  \odot \delta g  \odot \delta h  \odot \delta i)$,
$k( \delta^2 f \odot \delta^2 g)$,
$k( \delta^2 f \odot \delta g  \odot \delta h)$,
$k( \delta^3 f  \odot \delta g)$.

These eight differentials can be expressed in terms of the $2^4=16$ generators
$\delta_I$, with $I \subset \{3,2,1,0\}$, as follows

\begin{eqnarray*}
 \delta^4 f & = &  \delta_{3210}f \\
 \delta f \odot \delta^3 g  & = &  \delta_3 f \delta_{210}g \\
 \delta f \odot \delta^2 g \odot \delta h  & = & \delta_3 f (\delta_{21}g\delta_0h+\delta_1g\delta_{20}h-\delta_2g\delta_{10}h) \\
{} & = & \delta_3 f \delta_{21}g\delta_0h+ \delta_3 f \delta_1g\delta_{20}h-
\delta_3f \delta_2g\delta_{10}h \\
 \delta f \odot \delta g  \odot \delta^2 h  & = & \delta_3f (\delta_2g \delta_{10}h) = \delta_3f \delta_2g \delta_{10}h \\
 \delta f  \odot \delta g  \odot \delta h  \odot \delta i  & = & 
\delta_3 f \delta_2 g \delta_1 h \delta_0 i\\
 \delta^2 f \odot \delta^2 g  & = & \delta(\delta f \odot \delta^2 g)
- \delta f \odot \delta^3 g = \delta_3(\delta_2f\delta_{10}g)-\delta_3f\delta_{210}g\\
{} & = & \delta_{32}f \delta_{10}g + \delta_2 f \delta_{310}g -\delta_3f\delta_{210}g \\
 \delta^2 f \odot \delta g  \odot \delta h  & = & 
\delta(\delta f\odot\delta g\odot\delta h)-\delta f \odot \delta^2 g \odot
 \delta h - \delta f \odot \delta g \odot \delta^2 h \\
{} & = & \delta_3(\delta_2f\delta_1g\delta_0h)-
(\delta_3f\delta_{21}g\delta_0h+\delta_3f\delta_1g\delta_{20}h-\delta_3f\delta_2g\delta_{10}h)
 -\delta_3f\delta_2g\delta_{10}h \\
{} & = & \delta_{32}f\delta_1g\delta_0h+\delta_2f\delta_{31}g\delta_0h +
\delta_2f\delta_1g\delta_{30}h -\delta_3f\delta_{21}g\delta_0h \\
 {} & {} &  - \delta_3f\delta_1g\delta_{20}h + \delta_3f\delta_2g\delta_{10}h -
\delta_3f\delta_2g\delta_{10}h
\\
 \delta^3 f  \odot \delta g  & = & \delta(\delta^2 f \odot \delta g) - 
\delta^2f \odot \delta^2g \\
{} & = & \delta_3(\delta_{21}f\delta_0g+\delta_1f\delta_{20}g-\delta_2f\delta_{10}g)
-\delta_{32}f\delta_{10}g \\
{} & {} & -\delta_2f\delta_{310}g+\delta_3f\delta_{210}g \\
{} & = & 
\delta_{321}f\delta_0g + \delta_{21}f\delta_{30}g+\delta_{31}f\delta_{20}g+
\delta_1f\delta_{320}g \\
{} & {} & -\delta_{32}f\delta_{10}g-\delta_2f\delta_{310}g
-\delta_{32}f\delta_{10}g-\delta_2f\delta_{310}g+\delta_3f\delta_{210}g
\end{eqnarray*}
\end{description}
We see clearly on this example the interest of the Leibniz notation. For instance, the form of order $4$ equal to $\delta^3f \odot \delta g$ is quite a complicated object when written in terms of the differentials $\delta_I$ of the iterated frame algebras!
It is even worse (and at least uses a lot of space) if we write it explicitly in terms of
tensor products since it is an element of ${\cal A}^{\otimes 16}$.

\section{Representations of higher differentials. Examples}

In order to define differential forms of order $n$ we made use of some prior knowledge on the
algebras of universal forms $\Omega(F_p({\cal A}))$ associated with
appropriate iterated frame algebras, along with 
their differentials $\delta_p = d_{2^p - 1}$. We could have also used
homomorphic images of these algebras of universal forms, for instance, in the case
where ${\cal A}$ is commutative, we could have used
De Rham $p$-forms rather than functions of $p$ variables vanishing on
consecutive diagonals. We shall illustrate such a possibility below.

\subsection{A commutative example}

\subsubsection{Universal forms}

Take ${\cal A}=C(M)$, the algebra of continuous functions on a compact
topological space. Elements of $\Omega^1{\cal A}$ are linear combinations 
of elements of the kind $f\delta g = f\otimes g - fg \otimes 1$, so that they
can be identified with functions of two variables vanishing on the diagonal
, \ie $[f\delta g](x,y)=f(x)g(y) - f(x)g(x)=f(x)(g(y)-g(x)).$ We do not need to
use $\Omega^s{\cal A}$, $s >1$ , \ie universal forms of dergree $s$, in this paper, but let us mention, for
illustration purposes, that its elements 
can be identified with functions of $s+1$ variables vanishing on pairwise 
diagonals; for instance, elements of $\Omega^3{\cal A}$,
are linear combinations of functions of $4$ variables, of the kind
$$[f\delta g \delta h \delta k](x,y,z,t)=f(x)(g(y)-g(x))(h(z)-h(y))(k(t)-k(z)).$$

Let us compare this  with forms of order $s$, \ie  elements 
${\cal D}_s{\cal A}$. A priori, they are included in $F_s{\cal A}$ which is
itself isomorphic with $A_{2^s}$, so that these elements can be
considered as functions of $2^s$ variables over the space $M$.

 We already know that  
${\cal D}_1{\cal A} = \Omega_1{\cal A}$ so that forms of order $1$ are of the kind $$
f\delta g = f\otimes g - fg \otimes 1
$$
These forms, as it is well known, can be identified with functions of $2$ variables that vanish on the diagonal, since
$$
[f\delta g](x,y) = f(x)g(y) - f(x) g(x) = f(x) (g(y) - g(x))
$$

Forms of order $2$, \ie elements of ${\cal D}_2{\cal A}$
can be of the type $f\delta^2 g$, in which case
(see section 5.2)

\begin{eqnarray*}
f\delta^2 g & = & f\delta_1\delta_0 g \\ 
& = & f({\underline 1}_1 \otimes (1\otimes g - g\otimes 1)-
(1\otimes g - g\otimes 1)\otimes {\underline 1}_1) \\
{} & {} &
f\otimes 1 \otimes 1 \otimes g - f \otimes 1 \otimes g 
\otimes 1
- f\otimes g \otimes 1 \otimes 1 +
f g\otimes 1 \otimes 1 \otimes 1
\end{eqnarray*}
So that
\begin{eqnarray*}
[f\delta^2 g](x,y,z,t) & = &
f(x)g(t)-f(x)g(z) - f(x)g(y) + f(x)g(x)\\ {} & = &
f(x)((g(t)-g(z))-(g(y)-g(x))))
\end{eqnarray*}
Note that such forms can be identified with functions of $4$ variables $x,y,z,t$ that vanish when {\it both \/} $x=y$ {\it and \/}$z=t$
{\it or\/} when {\it both \/} $x=z$ {\it and \/} $y=t$.
.
Forms of order $2$
can also be of the type $f\delta g \odot \delta h$, in which case
(see section 5.2)

\begin{eqnarray*}
f\delta g \odot \delta h & = & f \delta_1 \rho_0 g \lambda_1 \delta_0 h \\
& = &
f \otimes 1 \otimes g \otimes h -
fg \otimes 1 \otimes 1 \otimes h -
f \otimes 1 \otimes gh \otimes 1 +
fg \otimes 1 \otimes h \otimes 1)
\end{eqnarray*}
So that
$$
[f\delta g \odot \delta h](x,y,z,t)=
f(x)(g(z)-g(x))(h(t)-h(z))
$$
Notice that such functions of $4$ variables $x,y,z,t$ vanish when {\it both} $x=z$ 
{\it and\/} $z=t$. 

\subsubsection{local forms}

We now suppose that $M$ is a compact smooth manifold,
take ${\cal A} = C^\infty(M)$, and replace the algebras of
 universal forms $\Omega {\cal A}$ by the usual De Rham complex $\Lambda M$,
\ie non local forms $[\delta f](x,y)=f(y)-f(x)$ (discrete differences) by
the usual differentials $df = {\partial f \over \partial x^\mu} dx^\mu$, 
where  $\{x^\mu\}$ is some local
coordinate system. These $x^\mu$ are themselves elements of ${\cal A}$.
We shall continue to use the notation $\delta f$.
 The differential
form of order $2$, that we called $\delta^2 f$ can, {\it a priori}, be expanded on the two types
of generators that span ${\cal D}_2{\cal A}$, \ie, 
$$\delta^2 f = a_{\mu \nu} \delta x^\mu \odot \delta x^\nu + b_\mu \delta^2 x^\mu$$
The form of order one $\delta f$ is equal to the  usual one-form $\delta_0 f$, therefore
$\delta f = (\partial_\mu f) \delta x^\mu$, where $\partial_\mu$ denotes the
coordinate frame $\partial_\mu \doteq {\partial \over {\partial x^\mu}}$,
so that $\delta^2 f= \delta(\partial_\mu f) \odot \delta x^\mu
+ (\partial_\mu f)\delta^2 x^\mu$. We have therefore to identify $b_\mu$ with the first derivatives $\partial_\mu f$ and
identify  $a_{\mu \nu}$ with the second derivatives $\partial_{\mu\nu}f$ of $f$
with respect to this coordinate frame.

Let us take for instance $ M = \RR^2$, and call $(x,y) \doteq (x^1,x^2)$. Then
$$
\delta^2 f = f_{xx}'' \delta x \odot \delta x + f_{yy}'' \delta y \odot \delta y +
 f_{xy}'' \delta x \odot \delta y +  f_{yx}'' \delta y \odot \delta x + f_x' \delta^2x
+f_y' \delta_2y
$$
We shall suppress the symbol $\odot$ in the remaining part of this section.
Using the property $f_{xy}'' = f_{yx}''$, we recover a result known to mathematicians of the
XIX century, namely that 
$$
\delta^2 f =  f_{xx}'' {\delta x}^2 + f_{yy}'' {\delta y}^2 + 2  f_{xy}''\delta x \delta y 
+ f_{x}' \delta^2 x + f_{y}' {\delta^2} y
$$
is an ``intrinsic'' quantity, hence invariant under a change of variables (coordinate system).
This remark can be used as follows. Let $x=x(u,v)$ and $y=y(u,v)$ a change of variables. We first
write
\begin{eqnarray*}
\delta x & = & x_u' \delta u + x_v' \delta v \\
\delta y & = & y_u' \delta u + y_v' \delta v \\
\delta^2 x & = &  x_{uu}'' {\delta u}^2 + x_{vv}'' {\delta v}^2 + 2  x_{uv}''\delta u \delta v 
+ x_{u}' \delta^2 u + x_{v}' {\delta^2} v \\
\delta^2 y & = &  y_{uu}'' {\delta u}^2 + y_{vv}'' {\delta v}^2 + 2  y_{uv}''\delta u \delta v 
+ y_{u}' \delta^2 u + y_{v}' {\delta^2} v 
\end{eqnarray*}
and replace these expressions by their values in $\delta^2 f$. We obtain
\begin{eqnarray*}
\delta^2f & = & (f_{xx}'' {x_u'}^2 + f_{yy}''  {y_u'}^2 + 2 f_{xy}'' {x_u'}{y_u'}+f_x' x_{uu}''
+f_y' y_{uu}''){\delta u  }^2  + \\
{} & {} &  (f_{xx}'' {x_v'}^2 + f_{yy}''  {y_v'}^2 + 2 f_{xy}'' {x_v'}{y_v'}+f_x' x_{vv}''
+f_y' y_{vv}''){\delta v  }^2  + \\
{} & {} & 2  (f_{xx}'' {x_u' } {x_v'}+ f_{yy}''  {y_u'} {y_v'} + f_{xy}''( {x_u'}{y_v'}+ {x_v'}{y_u'}
 +f_x' x_{uv}'' + f_y' y_{uv}'')) {\delta u} {\delta v}  + \\
{} & {} & {} (x_u' f_x' + y_u' f_y')\delta^2 u +  (x_v' f_x' + y_v' f_y')\delta^2 v
\end{eqnarray*}
Since $\delta^2 f$ is a ``geometrical quantitv'' it should be identified with
$$
\delta^2 f =  f_{uu}'' {\delta u}^2 + f_{vv}'' {\delta v}^2 + 2  f_{uv}''\delta u \delta v 
+ f_{u}' \delta^2 u + f_{v}' {\delta^2} v
$$
Therefore, we obtain
\begin{eqnarray*}
f_{uu}'' & = & f_{xx}'' {x_u'}^2 + f_{yy}''  {y_u'}^2 + 2 f_{xy}'' {x_u'}{y_u'}+f_x' x_{uu}''
+f_y' y_{uu}'' \\
f_{vv}'' & = &  f_{xx}'' {x_v'}^2 + f_{yy}''  {y_v'}^2 + 2 f_{xy}'' {x_v'}{y_v'}+f_x' x_{vv}''
+f_y' y_{vv}'' \\
f_{uv}'' & = & f_{xx}'' {x_u' } {x_v'}+ f_{yy}''  {y_u'} {y_v'} + f_{xy}''( {x_u'}{y_v'}+ {x_v'}{y_u'}
 +f_x' x_{uv}'' + f_y' y_{uv}'')  \\
f_{u}' & = & x_u' f_x' + y_u' f_y'  \\
f_{v}' & = & x_v' f_x' + y_v' f_y'
\end{eqnarray*}
Notice that, if the change of variables is linear, we can forget about the term 
$ f_{u}' \delta^2 u + f_{v}' {\delta^2} v$ in the expression of $\delta^2 f$
when we want to calculate the expression of $f_{..}''$, since, in this case,
$x_{..}''=y_{..}''=0$. However, for an arbitrary change of variables, this
term should be present! We see that the usual second order differential of $f$ is ``bad'' in the sense that
it presisely forgets the contribution of this term.
This fact (along with some lore concerning the definition of differentials of higher order) 
was well known to Bertrand or Hadamard (\cite{Bertrand}, \cite{Hadamard}, \cite{Drozbek} and reflects the fact that,
transforming between coordinate
systems requires that the second derivative be accompanied by the first$\ldots$
Instead of the chain rule for first derivatives, $d\phi/dv = d\phi/du \, du/dv$, we have something
more complicated for second derivatives, namely,
$d^2\phi/dv^2 = d^2\phi/du^2 \, (du/dv)^2 + d\phi/du \, d^2u/dv^2$.
These two rules of transformation for first {\it and\/} second derivatives
 compress into the following matrix equation:
$$
\left(
\begin{array}{ccc}
{d\phi \over dv} & {d^2\phi \over dv^2}
\end{array}
\right)
=
\left(
\begin{array}{ccc}
{d\phi \over du} & {d^2\phi \over du^2}
\end{array}
\right)
\left(
\begin{array}{ccc}
{du \over dv} & {d^2u\over dv^2} \\
0 & ({du \over dv})^2
\end{array}
\right)
$$
The interested reader may look at \cite{Foster} for an amusing ---and instructing---
elaboration along these lines.

\subsection{The example of algebra of $p\times p$ matrices}

Let ${\cal A}$ be the algebra of $p \times p$ matrices over the complex numbers.
For instance, take $p=2$. It is then  easy to construct the differentials of order $1, 2, 3 \ldots$ by using the explicit description of these objects in terms of tensor products and by 
performing tensor products of matrices.
Call $f^i_j$ the matrix elements of the $2\times 2$ matrix $f$.

The form of order one
$\delta f = 1 \otimes f - f \otimes 1 $, is the $4\times 4$ matrix
\begin{eqnarray*}
\delta f  &  =  &
\left(
\begin{array}{cc}
1 & 0 \\
0 & 1
\end{array}
\right)
 \otimes
 \left(
\begin{array}{cc}
f^1_1 & f^1_2 \\
f^2_1 & f^2_2
\end{array}
\right)
-
 \left(
\begin{array}{cc}
f^1_1 & f^1_2 \\
f^2_1 & f^2_2
\end{array}
\right)
\otimes
\left(
\begin{array}{cc}
1 & 0 \\
0 & 1
\end{array}
\right)
\\
&  =  &
 \left(
\begin{array}{cccc}
f^1_1 & 0 & f^1_2 & 0 \\
0 & f^1_1 & 0 & f^1_2 \\
f^2_1 & 0 & f^2_2 & 0 \\
0 & f^2_1 & 0 & f^2_2 
\end{array}
\right)
-
\left(
\begin{array}{cccc}
f^1_1 & f^1_2 & 0 & 0 \\
f^2_1 & f^2_2 & 0 & 0 \\
0 & 0 & f^1_1 & f^1_2 \\
0 & 0 & f^2_1 & f^2_2
\end{array}
\right)
=
\left(
\begin{array}{cccc}
0 & -f^1_2 & f^1_2 & 0 \\
-f^2_1 & f^1_1 - f^2_2 & 0 & f^1_2 \\
f^2_1 & 0 &f^2_2 -f^1_1 & -f^1_2 \\
0 & f^1_2 & -f^2_1 & 0
\end{array}
\right)
\end{eqnarray*}

 The form of order two
$\delta^2 f = \delta_1 \delta_0 f = {\underline 1}_1 \otimes \delta_0 f - \delta_0 f \otimes {\underline 1}_1 $, is the $16 \times 16$-matrix
\begin{eqnarray*}
\delta^2 f  &  =  &
\left(
\begin{array}{cccc}
1 & 0 & 0 & 0 \\
0 & 1 & 0 & 0 \\
0 & 0 & 1 & 0 \\
0 & 0 & 0 & 1
\end{array}
\right)
 \otimes
\left(
\begin{array}{cccc}
0 & -f^1_2 & f^1_2 & 0 \\
-f^2_1 & f^1_1 - f^2_2 & 0 & f^1_2 \\
f^2_1 & 0 &f^2_2 -f^1_1 & -f^1_2 \\
0 & f^1_2 & -f^2_1 & 0
\end{array}
\right)
\\
& {}  & -
\left(
\begin{array}{cccc}
0 & -f^1_2 & f^1_2 & 0 \\
-f^2_1 & f^1_1 - f^2_2 & 0 & f^1_2 \\
f^2_1 & 0 &f^2_2 -f^1_1 & -f^1_2 \\
0 & f^1_2 & -f^2_1 & 0
\end{array}
\right)
\otimes
\left(
\begin{array}{cccc}
1 & 0 & 0 & 0 \\
0 & 1 & 0 & 0 \\
0 & 0 & 1 & 0 \\
0 & 0 & 0 & 1
\end{array}
\right)
\\
&  =  & \ldots
\end{eqnarray*}

Such explicit calculations are very easy to handle with the help of a computer
but obviously require a lot of space when using only pen and paper!

The reader should carefully distinguish a form of degree $2$
 like $\delta_0 f \,  \delta_0 g$ (an element of $\Omega^2 {\cal A}$), which, in the present example is a $8 \times 8 $ matrix, from a Leibniz form of order $2$ like $\delta f \, \delta g \doteq \delta_1f \, \delta_0 g$ which, in the present example, is a $16 \times 16$ matrix.

\subsection{The algebra $\CC \oplus \CC$ of functions over two points}

Consider a discrete set $\{L,R\}$ with two elements that
we call $L$ and $R$. Call $x$ the coordinate function $x(L) \doteq 1, \, 
x(R) \doteq 0$ and  $y$ the coordinate function $y(L) \doteq 0,\,  y(R)
\doteq 1$. Notice that  $xy=yx=0$, $x^2=x, y^2=y$ and $x + y = 1$
where $1$ is the unit function $1(L)=1, 1(R)=1$. An arbitrary element
of the associative (and commutative) algebra ${\cal A}$ generated by
$x$ and $y$ can be written $\lambda x + \mu y$ (where $\lambda$ and
$\mu$ are two complex numbers) and can be represented as a diagonal
matrix $\pmatrix{\lambda & 0 \cr 0 & \mu \cr}$. One can write ${\cal
A} = \CC x \oplus \CC y$ and is isomorphic with $\CC \oplus \CC$.

 We now introduce a differential $\delta_0$ satisfying $\delta_0^2=0$,
$\delta_0 1 = 0$ and the usual Leibniz rule, along with formal symbols
$\delta_0 x$ and $\delta_0 y$. It is clear that $\Omega^1$, the space of
differentials of degree $1$ is generated by the two independent
quantities $x\delta_0 x$ and $y\delta_0 y$. Indeed, the relation $x+y=1$
implies $\delta_0 x + \delta_0 y = 0$, the relations $x^2=x$ and $y^2=y$
imply $(\delta_0 x)x+x(\delta_0 x) = (\delta_0 x)$, therefore $(\delta_0 x) x
= (1-x) \delta_0 x$ and $(\delta_0 y) y = (1-y) \delta_0 y$. This implies
also, for example $\delta_0 x = 1 \delta_0 x = x \delta_0 x + y \delta_0 x$,
$x \delta_0 x = - x \delta_0 y$, $y \delta_0 x = (1-x) \delta_0 x$, $(\delta_0
x)x = y\delta_0 x = - y \delta_0 y$ \etc.

 More generally,
$\Omega^p$, the space of differentials of degree $p$ is also $2$-dimensional.
 The above
relations indeed imply that a base of this vector space is given by
$\{x\delta_0 x\delta_0 x\ldots\delta_0 x,y\delta_0 y\delta_0 y\ldots\delta_0 y 
\}$.

As we already know, the space $\Omega = \bigoplus_p \Omega^p$
 is an algebra: we multiply forms freely
but take into account the Leibniz rule.
Notice that $x \delta_0 x = x\otimes x - x^2 \otimes 1 = 
 x\otimes x - x \otimes 1$ since $x^2 = x$.
In the same way
$y \delta_0 y = y\otimes y - y \otimes 1$.
 Therefore
\begin{eqnarray*}
\lbrack x \delta_0 x \rbrack(L,L) = 0 &
 \mbox{and} & \lbrack y \delta_0 y \rbrack(L,L) = 0 \\
\lbrack x \delta_0 x \rbrack (L,R) = -1 & \mbox{and} & \lbrack y \delta_0 y \rbrack (L,R) = 0 \\
\lbrack x \delta_0 x \rbrack (R,L) = 0 & \mbox{and} & \lbrack y \delta_0 y \rbrack (R,L) = -1 \\
\lbrack x \delta_0 x \rbrack (R,R) = 0 & \mbox{and} & \lbrack y \delta_0 y \rbrack (R,R) = 0 
\end{eqnarray*}
We knew, {\it priori\/} that these two functions had to vanish on the
diagonal, \ie on the arguments $(L,L)$ and $(R,R)$.

Our main interest, in this paper, is not in the study of 
 spaces $\Omega^p{\cal A}$, with $p>1$ but in the spaces ${\cal D}_q{\cal A}$.
Using the fact that  ${\cal D}_2{\cal A}$ is spanned by forms of the type $f\delta^2 g$ or of the type $f\delta g \odot \delta h$, the reader will easily show that
 ${\cal D}_2{\cal A}$ is spanned by the four monomials $x \delta^2 x$, 
$y \delta^2 y$, $x \delta x \delta x$ and $y \delta y  \delta y$
(again we do not write explicitly the symbol $\odot$). Here $\delta^2 x \doteq \delta_1 \delta_0 x$ where $\delta_1$ is the differential in $\Omega^1{\cal A}\otimes {\cal A})$. We also had to use relations such as
 $\delta^2 x = - \delta^2 y$, $x \delta^2 x + y \delta^2 x = \delta^2 x $, \etc coming from the relations in the algebra ${\cal A}$.

Explicitly, one obtains, for instance (use $x^2 = x, y^2 = y, xy=0$)
$$
x \delta^2 x = x \otimes 1  \otimes 1  \otimes x - 
x \otimes 1  \otimes x  \otimes 1 -
x  \otimes x  \otimes 1  \otimes 1 +
x  \otimes 1  \otimes 1  \otimes 1
$$ 
and
$$
x \delta x \delta x =
x  \otimes 1  \otimes x  \otimes x -
x \otimes 1  \otimes 1  \otimes x$$
Note that the last two terms appearing in the general expression of $f\delta g \delta h$ cancel each other.

 For illustration, the reader may convince himself
that the only non-zero values
 of $\lbrack x \delta^2 x\rbrack (x_1,x_2,x_3,x_4)$ are the following ones:
\begin{eqnarray*}
  \lbrack x \delta^2 x\rbrack (L,L,L,R) = -1,
& \lbrack x \delta^2 x\rbrack (L,L,R,L) =  1,
& \lbrack x \delta^2 x\rbrack (L,R,L,L) = -1, \\
  \lbrack x \delta^2 x\rbrack (L,R,R,R) = -1,
& \lbrack x \delta^2 x\rbrack (L,R,R,L) =  2,
& {}
\end{eqnarray*}
In the same way, one can show that ${\cal D}_3{\cal A}$ is spanned by the eight types $8 = 2^{(3-1)}\times 2$ of monomials
\begin{eqnarray*}
 x\delta^3x,\, y \delta^3 y  \,(\mbox{with a structure}  f\delta^3 g ),& {} &
 x\delta^2 \delta x,\, y \delta^2 y \delta y  \,(\mbox{with a structure}  f\delta^2 g \delta h ),
\\
 x\delta x\delta^2 x,\, y \delta y\delta^2 y  \,(\mbox{with a structure}  f\delta g\delta^2 h ), & {} &
 x(\delta x)^3,\, y (\delta y)^3  \,(\mbox{with a structure}  f\delta g\delta h \delta u )
\end{eqnarray*}
Since elements of ${\cal A}$ can be represented by diagonal $2\times 2$ matrices,
this last example can be considered as a particular case of the previous one.
Setting $f = diag(\lambda , \mu)$,  and $\epsilon = \mu - \lambda$, we obtain
for instance, $\delta f = diag(0,-\epsilon,\epsilon,0)$ and
$$\delta^2 f = diag(0,-\epsilon,\epsilon,0,-\epsilon,-2\epsilon,0,-\epsilon,
\epsilon,0,2\epsilon,\epsilon,0,-\epsilon,\epsilon,0)
$$
\bigskip

{\Large Acknowledgments}
\bigskip

Part of this work was done during my stay at the University of Zaragoza, in April 1996 and I would like to thank members from the theoretical physics department of this university, in particular Prof. M. Asorey, for the hospitality and for providing a friendly atmosphere.
I would also like to thank Mrs T. Stavracou, at CPT, for her comments.

\eject

\end{document}